\documentclass{elsarticle}

\usepackage{amssymb}

\usepackage[figuresright]{rotating}

\begin{document}

\begin{frontmatter}




\title{From ANAIS-25 towards ANAIS-250}


\author[1,2]{J.\,Amar\'e }
\author[1,2]{S.\,Cebri\'an}
\author[1,2]{C.\,Cuesta\footnote{Present Address: Center for Experimental Nuclear Physics and Astrophysics and Department of Physics, University of Washington, WA, US}}
\author[1,2]{E.\,Garc\'{\i}a}
\author[1,2]{C.\,Ginestra}
\author[1,2,3]{M.\,Mart\'{\i}nez}
\author[1,2]{M.A.\,Oliv\'an}
\author[1,2]{Y.\,Ortigoza}
\author[1,2]{A.\,Ortiz de Sol\'orzano}
\author[1,2]{C.\,Pobes\footnote{Present Address: Instituto de Ciencia de Materiales de Arag\'on, Universidad de Zaragoza - CSIC, Zaragoza, Spain}}
\author[1,2]{J.\,Puimed\'on}
\author[1,2]{M.L.\,Sarsa\footnote{Corresponding author, mlsarsa@unizar.es}}
\author[1,2]{J.A.\,Villar}
\author[1,2]{P.\,Villar }

\address[1]{Laboratorio de F\'{\i}sica Nuclear y Astropart\'{\i}culas, Universidad de Zaragoza, C/Pedro Cerbuna 12, 50009 Zaragoza, SPAIN}
\address[2]{Canfranc Underground Laboratory, Paseo de los Ayerbe s.n., Canfranc Estaci\'on, Huesca, SPAIN}
\address[3]{Fundaci\'on ARAID, Mar\'{\i}a de Luna 11, CEEI Arag\'on, 50018 Zaragoza, SPAIN}

\begin{abstract}
The ANAIS (Annual modulation with NaI(Tl) Scintillators) experiment aims at the confirmation of the DAMA/LIBRA signal using the same target and technique at the Canfranc Underground Laboratory (LSC). 250\,kg of ultra pure NaI(Tl) crystals will be used as target, divided into 20 modules, 12.5\,kg mass each, and coupled to two high efficiency photomultiplier tubes from Hamamatsu. The ANAIS-25 set-up at the LSC consists of two prototypes, amounting 25\,kg NaI(Tl), grown from a powder having a potassium level under the limit of our analytical techniques, and installed in a convenient shielding at the LSC. The background has been carefully analyzed and main results will be summarized in this paper, focusing on the alpha contamination identified in the prototypes and the related background contributions. Status of fulfillment of ANAIS experimental goals and prospects for the building of ANAIS-250 experiment will be also revised. 
\end{abstract}


\end{frontmatter}


\section{Introduction}
\label{sec:intro}

ANAIS project aims at the study of the annual modulation signal attributed to galactic dark matter particles\,\cite{annual_modulation} using 250\,kg NaI(Tl) at the Canfranc Underground Laboratory (LSC), in Spain. The DAMA experiment, at the Laboratori Nazionali del Gran Sasso, in Italy, reported first evidence of the presence of an annual modulation in the detection rate compatible with that expected for a dark matter signal, just in the region below 6\,keVee (electron equivalent energy) with a high statistical significance\,\cite{DAMA}. This signal was further confirmed by the LIBRA experiment, using 250\,kg of more radiopure NaI(Tl) detectors\,\cite{LIBRA}. Using the same target than DAMA/LIBRA experiment makes possible for ANAIS to confirm such a result in a model independent way. To achieve this goal, ANAIS detectors should be as good (or better) as (than) those of DAMA/LIBRA in terms of energy threshold and radioactive background below 10\,keVee: energy threshold below 2\,keVee and background at 1-2\,counts/keV/kg/day.

After the operation of several prototypes at the LSC\,\cite{ANAIS_prototypes, ANAISbkg, ANAIS_RICAP}, the main challenge for ANAIS is the achievement of the required low background level, being contaminations in the bulk of the crystal still dominant in the background. However, backgrounds at low, medium and high energy are quite well understood, as shown with ANAIS-0 prototype\,\cite{ANAISbkg} and some other interesting results, as very slow scintillation in NaI(Tl)\,\cite{ANAISom} or an anomalous fast event population attributable to quartz scintillation\,\cite{ANAISquartz} have been obtained. In the following, we will report on the main results derived from the ANAIS-25 prototypes concerning background level and analysis of the dominant contributions (Section\,\ref{sec:bck}), light collection efficiency and energy threshold (Section\,\ref{sec:yield}). Before, we will briefly describe the experimental setup (Section\,\ref{sec:setup}) and, finally, prospects for the building of ANAIS-250 will be revised (Section\,\ref{sec:prospects}).


\section{ANAIS-25 experimental set-up}
 \label{sec:setup}

\begin {figure}[h!]
\centering{\includegraphics[width=0.65\textwidth]{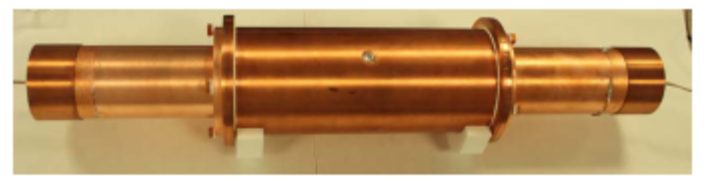}}
\caption{ANAIS-25:D0 module: 12.5\,kg mass crystal by Alpha Spectra coupled to two PMTs (Hamamatsu R12669SEL2) at the LSC clean room. Calibration window can be noticed.}
\label{fig:ANAIS-25module}
\end {figure}

The ANAIS-25 set-up consists of two cylindrical 12.5\,kg mass NaI(Tl) crystals grown by Alpha Spectra (AS) \cite{AlphaSpectra}  using ultra-pure NaI powder (below 90\,ppb potassium at 95\% C.L. according to HP Ge spectrometry results at the LSC). Detectors were manufactured by AS in their ultra-low background assembly laboratory. The University of Zaragoza (UZ) group collaborated with AS in the detector design, material selection and protocols for cleaning materials and tools. OFHC copper was used for the encapsulation with two synthetic quartz windows allowing the coupling of the photomultiplier tubes (PMTs) in a second step at the LSC clean room. An aluminized Mylar window allows to calibrate at low energy both detectors (see Figure~\ref{fig:ANAIS-25module}). 

Two units of R12669SEL2 and two units of R11065SEL Hamamatsu PMTs were coupled  to modules ANAIS-25:D0 and D1, respectively and the detectors were installed at LSC to start immediately the data taking inside a shielding consisting of: 10\,cm archaeological lead, 20\,cm low activity lead, PVC box tightly closed and continuously flushed with boil-off nitrogen gas, and active plastic scintillator vetoes placed on top of the shielding.

\section{Main background contributions in ANAIS-25}
 \label{sec:bck}
 
The main goal of the ANAIS-25 set-up was to determine the potassium content of the AS crystals. For that, the study of the coincidences between the two modules was used and a result for the $^{40}$K activity of 1.25\,$\pm$\,0.11\,mBq/kg (equivalent to 41.7\,$\pm$\,3.7\,ppb natural potassium) was derived \cite{ANAIS_40K}. AS is improving the NaI powder purification procedures in order to get down to the 20\,ppb potassium goal. 

\begin {figure}[h!]
\centering{\includegraphics[width=0.51\textwidth]{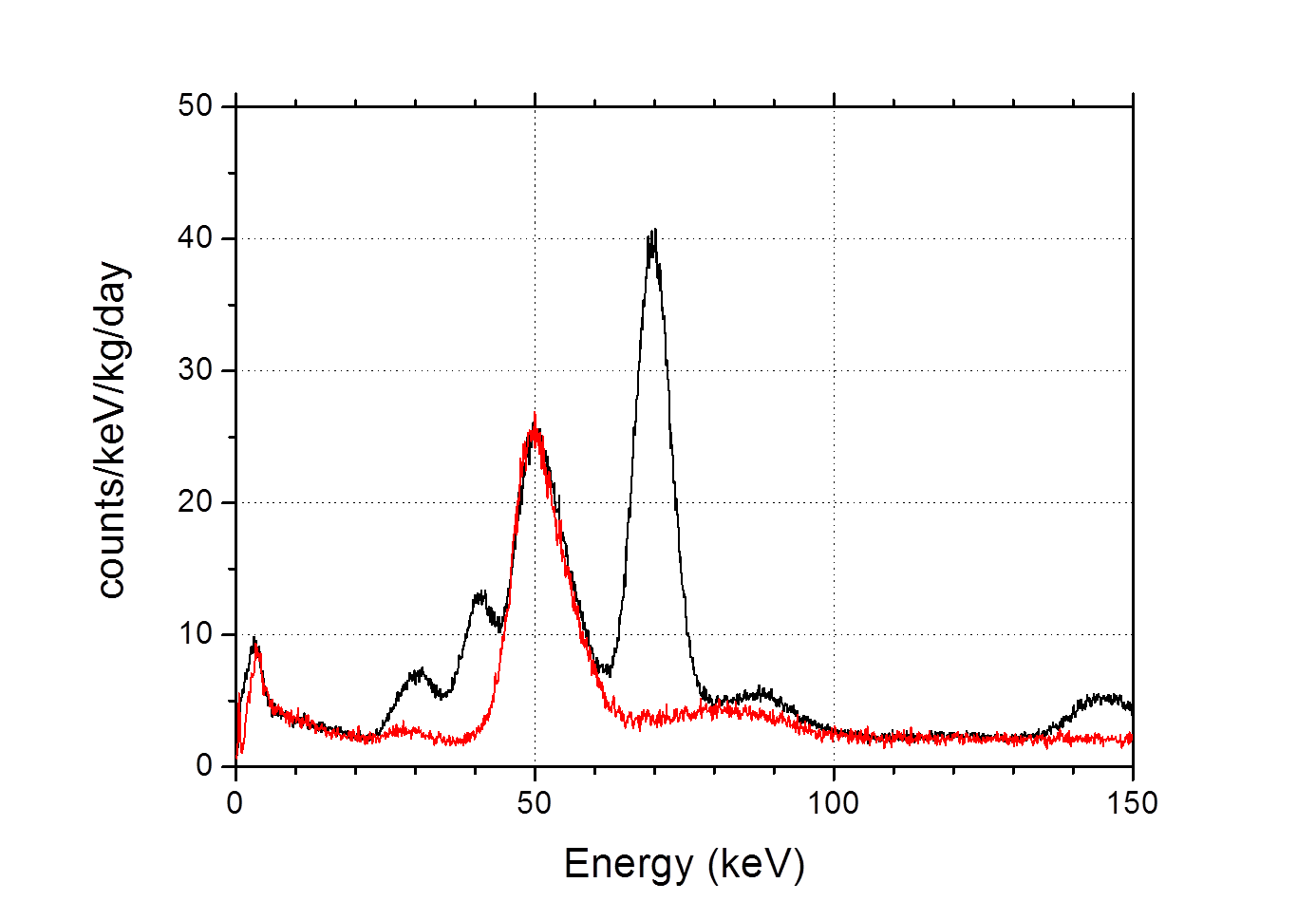}
\hspace{0.2cm} \includegraphics[width=0.45\textwidth]{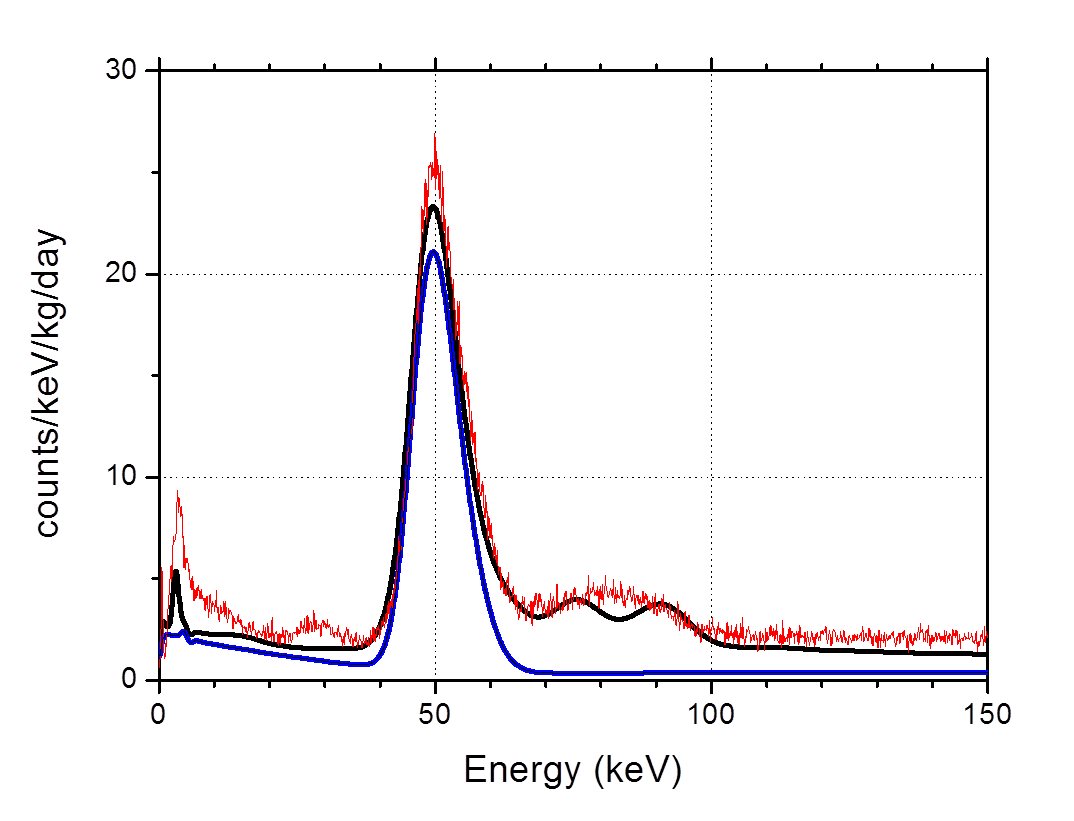}}
\caption{Left: Background at low energy from ANAIS-25:D0 module: at the beginning of the data taking (black), and after fifteen months of underground operation (red). Right: The latter of the previous backgrounds (red) compared to the simulation of the contribution from the different isotopes identified in the scintillator bulk, other than cosmogenic origin ones, (black) and the contribution of $^{210}$Pb in the crystal bulk (blue) at the level required to explain the total alpha rate measured. Contaminations from PMTs and other detector components determined by HPGe Spectrometry have been also included in the simulations. Filtering of events at the very low energy region has not been efficiency corrected.}
\label{fig:bkgLE}
\end {figure}

\begin {figure}[h!]
\centering{\includegraphics[width=0.65\textwidth]{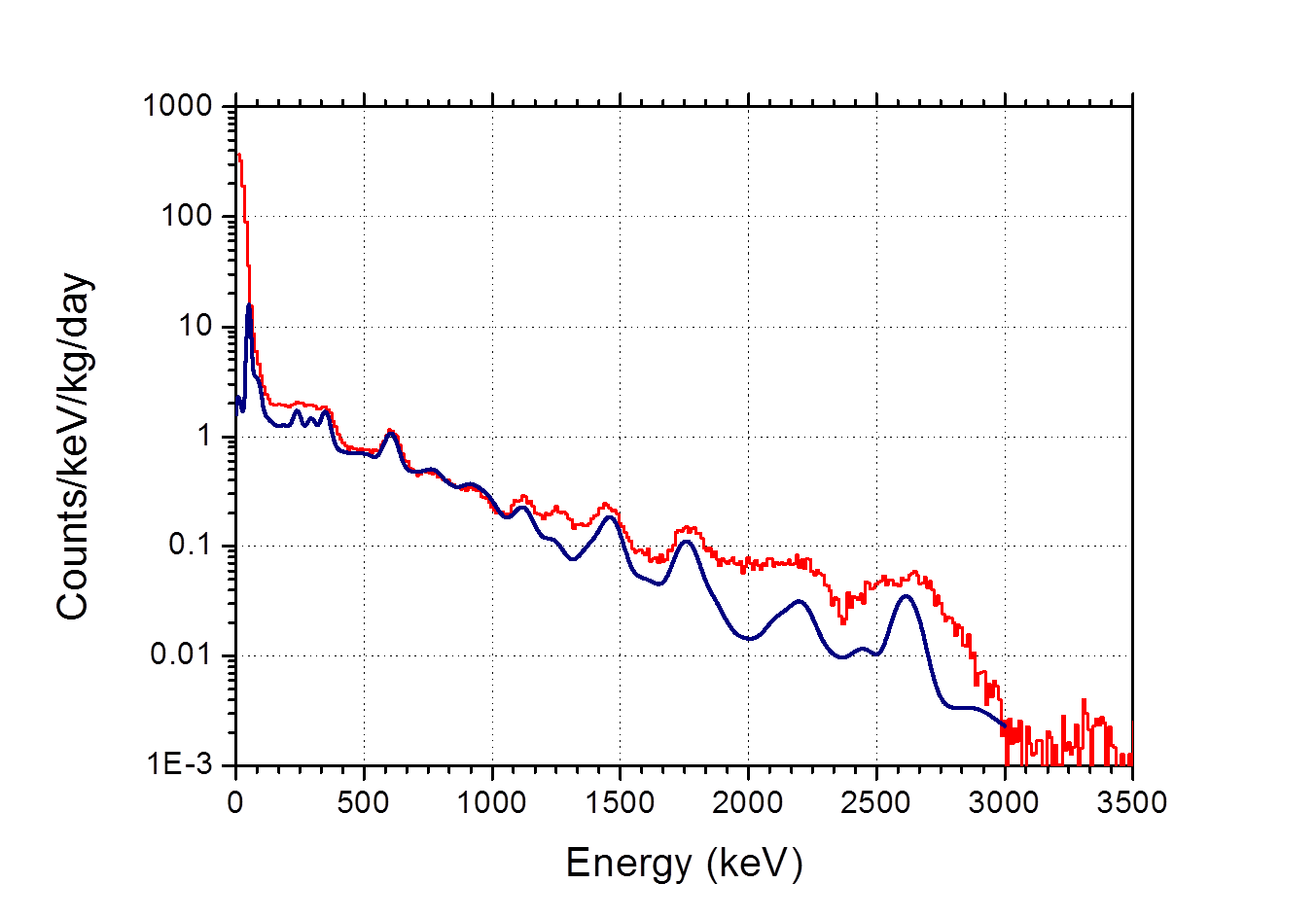}}
\caption{Gamma background at high energy from ANAIS-25:D0 module fifteen months after going underground (red) and corresponding simulation (blue). Contaminations in the NaI crystal bulk, as well as those from PMTs and other detector components determined by HPGe Spectrometry have been included in the simulation. }
\label{fig:bkgHE}
\end {figure}

We show in Figures\,\ref{fig:bkgLE} and \ref{fig:bkgHE} the background at low and high energy, respectively, for ANAIS-25:D0 module. In the latter, only gamma events are shown. Similar background is obtained for ANAIS-25:D1 module, although slightly lower at medium and high energy because of the PMTs used in the latter (Ham. R11065SEL model), which have better radiopurity. Cosmogenically activated isotopes contribution can be clearly identified in the first months of measurement underground\,\cite{ANAIS_RICAP}. 

Contaminations in the NaI crystal bulk, as well as those from PMTs and other detector components determined by HPGe Spectrometry have been included in the simulation using G\'eant4 package and following a similar approach to that presented in\,\cite{ANAISbkg}. In particular, in the crystal bulk have been simulated $^{40}$K$, ^{129}$I, and the isotopes from the $^{238}$U and $^{232}$Th chains identified as explained below; in  the case of $^{210}$Pb, it has been assumed that all the contamination required to explain the total alpha rate measured is found in the crystal bulk (see discussion below). Cosmogenic isotopes have not been included in the simulation, whose results are shown in Figures \,\ref{fig:bkgLE} and \ref{fig:bkgHE}.

Pulse Shape Analysis (PSA) allows to powerfully discriminate the alpha origin energy depositions in the bulk and, hence, to identify and quantify the presence of different isotopes from the $^{238}$U and $^{232}$Th chains. ANAIS-25 modules showed a total alpha rate: 3.15\,mBq/kg, further beyond the radiopurity goals of ANAIS experiment and the corresponding energy spectrum does not show typical features from natural chains. Efforts to understand the origin of the contamination are still ongoing, but some progress in this direction can be reported: the activities of some of the alpha-decaying isotopes of the natural chains present in ANAIS-25 modules have been identified through the $\alpha - \alpha$ and Bi-Po coincidences and are given in Table\,\ref{tab:chains}; it should be noticed the reduced activity in these isotopes compared to the level measured in previous prototype ANAIS-0. Because of that, it is difficult to explain the alpha rate observed in ANAIS-25 without strongly breaking equilibrium in at least one of the natural chains: the most plausible hypothesis is assuming $^{210}$Po as responsible of most of the alpha rate observed in ANAIS-25 set-up, not being possible to discard by now neither bulk, nor superficial contamination, nor even a combination of both. $^{210}$Po should have entered into the purification/growing/building process through progenitor $^{222}$Rn, which having a very short lifetime would have very fast decayed, implanting  $^{210}$Pb in the powder or crystal. Our low energy spectrum shows clearly the presence of $^{210}$Pb decay, being the observed events rate compatible with expectations for a bulk contamination at the level required to explain the whole alpha rate measured (see Figure\,\ref{fig:bkgLE}).  

\begin{table}[ht]
\begin{center}
\caption{Results for the activities of some of the isotopes in the natural chains in ANAIS-25 modules are compared to those identified in ANAIS-0 prototype. Activities derived from $\alpha-\alpha$ and  Bi-Po coincidences in ANAIS-25 have been assigned to the long-life nearest progenitor.}
\vspace{0.5cm}
{\begin{tabular}{@{}ccc@{}}
\hline
Isotope  & \multicolumn{2}{c}{Activity (mBq/kg)}\\ 	
  &  	ANAIS-0  & ANAIS-25\\
\hline
$^{226}$Ra & $ 0.098 \pm 0.004$  & $0.010 \pm 0.002 $  \\
$^{228}$Th  & $0.035 \pm 0.003 $ & $0.003 \pm 0.001$ \\
\hline
\end{tabular}
\label{tab:chains}}
\end{center}
\end{table}

\begin {figure}[h!]
\centering{\includegraphics[width=0.6\textwidth]{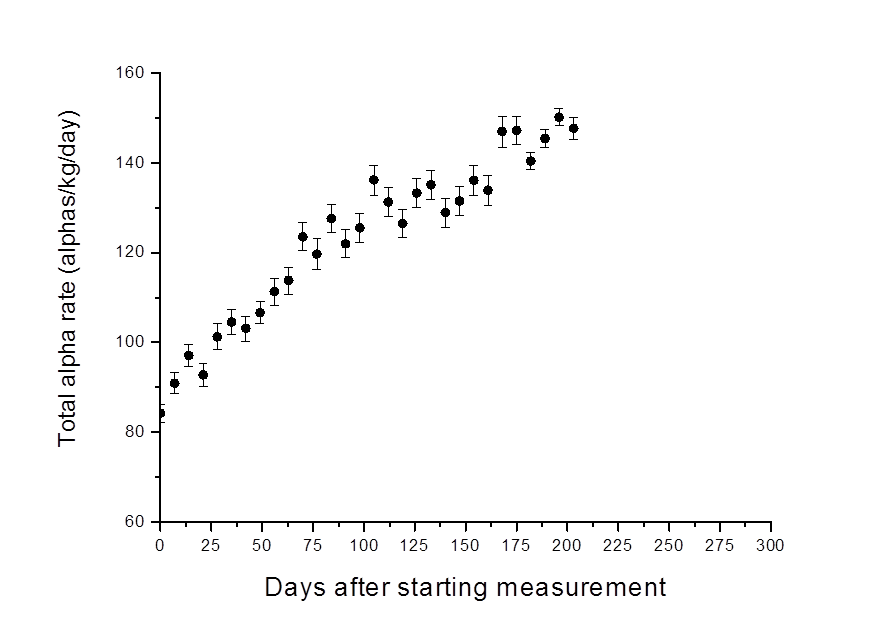}}
\caption{Total alpha rate observed in a 1\,kg mass crystal grown by Alpha Spectra and encapsulated by the UZ group. Data taking was carried out at LSC. Alpha rate increase is due to $^{210}$Po accumulating in the crystal after the decay of the progenitor, $^{210}$Pb.}
\label{fig:alpharate}
\end {figure}

We have confirmed this point by measuring at the LSC a new crystal, 1\,kg mass, grown by AS and sent without encapsulation to UZ. It was encapsulated and coupled to two PMTs at the UZ and installed for measuring at the LSC in a reduced shielding near ANAIS-25, having as only goal the determination of the alpha contamination. In this case, secular equilibrium had not yet been reached and a clear increase in the alpha rate compatible with the half-life of $^{210}$Po can be observed in Figure\,\ref{fig:alpharate}. Having the timing information of the procedures followed, mainly purification, handling, and growing, the origin of the contamination can be traced back. AS has modified, according to this information, its procedures to reduce the $^{210}$Pb content in the next crystals produced.

 \section{ANAIS-25 light collection efficiency}
 \label{sec:yield}
 
Total light collected per unit of energy deposited in the NaI(Tl) crystal has been calculated with the mean pulse area which corresponds to 22.6\,keV events from a $^{109}$Cd source and the mean pulse area of single photoelectrons, identified in pulses with a few number of them and averaged to build the Single Electron Response (S.E.R.). Corresponding efficiencies derived for ANAIS-0 prototype, similar in shape to DAMA/LIBRA crystals, using the same PMTs than ANAIS-25 modules, are also shown in Table\,\ref{tab:yield} for comparison. A remarkable improvement in light collection efficiency in the AS modules can be reported, pointing at the possibility of further reduction of the ANAIS threshold, demonstrated to be at 2\,keVee level in the ANAIS-0 prototype\,\cite{ANAISfiltrado}.

 \begin{table}[ht]
\begin{center}
\caption{Results for the effective amount of light collected per unit of energy deposited in ANAIS-25 modules compared to that of the previous prototype, ANAIS-0. Results using two different PMT models are shown.}
\vspace{0.5cm}
{\begin{tabular}{@{}ccc@{}}
\hline
PMT model  & ANAIS-0 & ANAIS-25\\ 
 & \multicolumn{2}{c}{phe/keV}\\		
\hline
Ham R12669SEL2 &  $7.38 \pm 0.07$  &  $16.13 \pm 0.66$\\
Ham R11065SEL\hphantom{0}   &   $5.34\pm 0.05 $  &  $12.58 \pm 0.13 $\\
\hline
\end{tabular}
\label{tab:yield}}
\end{center}
\end{table}

 \section{Prospects for ANAIS-250}
 \label{sec:prospects}
 
\begin {figure}[h!]
\centering{\includegraphics[width=0.75\textwidth]{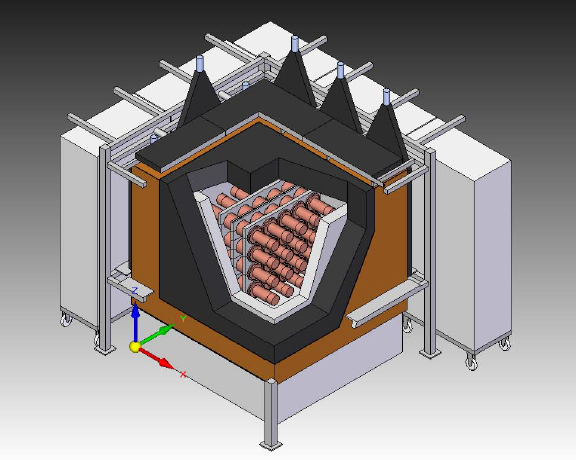} \vspace{0.3cm}}
\caption{Artistic drawing of the ANAIS250 setup in a 5x4 configuration, with 12.5\,kg mass cylindrical modules.}
\label{fig:ANAIS250}
\end {figure}

While crystal radiopurity goal is being achieved, the rest of experimental requirements for ANAIS experiment have been or are being tested at LSC. Summarizing the present status: 

\begin{itemize}
\item Light collection efficiency features and other related parameters of the ANAIS-25 modules have proven to be excellent (see previous section): remarkable optical quality of the crystals plus high efficient Hamamatsu PMTs contribute to this point.
\item Very strong tools to discriminate scintillation events steming from the NaI(Tl) bulk crystal from other kind of spurious events, having mainly the origin in the PMTs, have been developed.
\end{itemize}
These two features make us confident to lower ANAIS threshold beyond the  2\,keVee already achieved with previous prototypes, because total light collection efficiency more than doubled (see Table\,\ref{tab:yield}). At the same time, the improved energy resolution observed at very low energy in ANAIS-25 modules helps to reduce the possible interference of the $^{40}$K contribution in the dark matter analysis.
\begin{itemize}
\item The LSC experimental space for ANAIS is ready for the mounting: shielding materials are stored underground; 40 units of the Hamamatsu PMT model chosen for ANAIS have been ordered, and 15 units have been already screened for radiopurity at the LSC HP Ge test bench; plastic scintillators vetoes are being tested; and electronic chain and acquisition software have been fully commissioned and tested with the two modules set-up ANAIS-25.
\item A successful background model has been developed for ANAIS-0 prototype\,\cite{ANAISbkg} and is being applied to ANAIS-25 (some of the preliminary results have been shown in Section\,\ref{sec:bck}).
\item A full simulation of ANAIS-250 set-up consisting of 20 modules 12.5 kg mass each, cylindrical in shape, and installed in a 4x5 configuration, is ongoing (see Figure\,\ref{fig:ANAIS250}). One of the first results under analysis is the capability of rejecting $^{40}$K events as a function of different experimental parameters: copper encapsulation thickness, modules separation, modules total size, number of modules, etc.
\end{itemize}
 
 \section*{Acknowledgments}
This work has been supported by the Spanish Ministerio de Econom\'{\i}a y Competitividad and the European Regional Development Fund (MINECO-FEDER) (FPA2011-23749), the Consolider-Ingenio 2010 Programme under grants MULTIDARK CSD2009- 00064 and CPAN CSD2007-00042, and the Gobierno de Arag\'{o}n (Group in Nuclear and Astroparticle Physics, ARAID Foundation and C. Cuesta predoctoral grant). P.\,Villar is supported by the MINECO Subprograma de Formaci\'{o}n de Personal Investigador. We also acknowledge LSC and GIFNA staff for their support.





\end{document}